\documentclass[conference]{IEEEtran}
\usepackage[utf8]{inputenc}
\usepackage[T1]{fontenc}
\usepackage{cite}
\usepackage{booktabs}
\usepackage{url}
\newcommand{\cbTasks}{20}
\newcommand{\cbTauto}{13}
\newcommand{\cbMissed}{2}
\newcommand{\cbInformative}{7}
\newcommand{\cbCommit}{45cb1bf}
\newcommand{\calProjects}{5}
\newcommand{\calTruthCommit}{ce37eb5}
\newcommand{\pjSecFindings}{15}
\newcommand{\pjSecChain}{2}
\newcommand{\pjSecNegBefore}{7}
\newcommand{\pjSecDegraded}{2}
\newcommand{\pjSecNegAfter}{5}
\newcommand{\pjSecEntry}{6}
\newcommand{\pjTwoFindings}{6}
\newcommand{\pjTwoChain}{1}
\newcommand{\pjTwoNegBefore}{5}
\newcommand{\pjTwoDegraded}{0}
\newcommand{\pjTwoNegAfter}{5}
\newcommand{\pjTwoEntry}{0}
\newcommand{\ruleGlobalFloor}{0.05}
\newcommand{\ruleLocalSparse}{0.5}
\newcommand{\repPycgVerdicts}{51}
\newcommand{\repPycgDegradations}{29}
\newcommand{\repCbVerdicts}{20}
\newcommand{\popAsciinema}{134}
\newcommand{\popSublister}{98}
\newcommand{\appFaceClsNeg}{23}
\newcommand{\libTotalNeg}{56}
\newcommand{\libcTotalNeg}{25}
\newcommand{\fabAppManualReached}{1}
\newcommand{\fabAppPop}{81}
\newcommand{\cbLibcNeg}{5\,322}
\newcommand{\cbLibcDeg}{5\,322}
\newcommand{\survRefWrong}{2}
\newcommand{\sensLsrMin}{0.25}
\newcommand{\sensLsrMax}{1.0}
\newcommand{\sensGefMin}{0.01}
\newcommand{\sensGefMax}{0.1}
\newcommand{\sensAsciiDiffMin}{$-$16.4}
\newcommand{\sensAsciiDiffMax}{5.7}
\newcommand{\dynAsciinemaIntExec}{97}
\newcommand{\dynAsciinemaIntNegExec}{86}
\newcommand{\dynAsciinemaIntWrongExec}{39}
\newcommand{\dynAsciinemaIntAgreeExec}{47}
\newcommand{\dynAsciinemaIntRefNegExec}{47}
\newcommand{\dynAsciinemaExtExec}{119}
\newcommand{\dynAsciinemaExtNegExec}{108}
\newcommand{\dynAsciinemaExtWrongExec}{55}
\newcommand{\dynAsciinemaExtAgreeExec}{53}
\newcommand{\dynAsciinemaExtRefNegExec}{53}
\newcommand{\dynAsciinemaAllExec}{124}
\newcommand{\dynAsciinemaAllNegExec}{113}
\newcommand{\dynAsciinemaNeg}{123}
\newcommand{\dynAsciinemaAllWrongExec}{56}
\newcommand{\dynAsciinemaWrong}{57}
\newcommand{\dynAsciinemaAllAgreeExec}{57}
\newcommand{\dynAsciinemaAgree}{66}
\newcommand{\dynAsciinemaAllRefNegExec}{57}
\newcommand{\dynAsciinemaRefNeg}{66}
\newcommand{\dynAsciinemaPos}{11}
\newcommand{\dynAsciinemaAllRuleDeg}{62}
\newcommand{\dynAsciinemaAllRuleKept}{51}
\newcommand{\dynAsciinemaAllNotExecNeg}{10}
\newcommand{\dynAsciinemaAllNotExecDeg}{3}
\newcommand{\dynSublisterAllExec}{70}
\newcommand{\dynSublisterAllNegExec}{19}
\newcommand{\dynSublisterNeg}{24}
\newcommand{\dynSublisterAllWrongExec}{17}
\newcommand{\dynSublisterWrong}{18}
\newcommand{\dynSublisterAllAgreeExec}{2}
\newcommand{\dynSublisterAgree}{6}
\newcommand{\dynSublisterAllJoernOnlyExec}{41}
\newcommand{\dynSublisterJoernOnly}{58}
\newcommand{\dynSublisterAllRefNegExec}{43}
\newcommand{\dynSublisterRefNeg}{64}
\newcommand{\dynSublisterAllRuleDeg}{15}
\newcommand{\dynSublisterAllRuleKept}{4}
\newcommand{\dynSublisterAllNotExecNeg}{5}
\newcommand{\dynSublisterAllNotExecDeg}{2}
\newcommand{\dynSecTotal}{5}
\newcommand{\tcAsciinemaNegObserved}{113}
\newcommand{\tcAsciinemaMiss}{73}
\newcommand{\tcAsciinemaKindConstructor}{7}
\newcommand{\tcAsciinemaKindWith}{16}
\newcommand{\tcAsciinemaKindProperty}{13}
\newcommand{\tcAsciinemaKindClassValue}{7}
\newcommand{\tcAsciinemaKindAttribute}{18}
\newcommand{\tcAsciinemaKindParamValue}{1}
\newcommand{\tcAsciinemaKindSelfOverride}{0}
\newcommand{\tcAsciinemaKindSelfMethod}{0}
\newcommand{\tcAsciinemaKindViaLibrary}{4}
\newcommand{\tcAsciinemaKindGenerator}{7}
\newcommand{\tcAsciinemaLeftDispatch}{46}
\newcommand{\tcAsciinemaLeftImplicit}{103}
\newcommand{\tcAsciinemaLeftDispatchImplicit}{12}
\newcommand{\tcAsciinemaAloneDispatch}{67}
\newcommand{\tcAsciinemaAloneImplicit}{10}
\newcommand{\tcAsciinemaOnlyJointly}{35}
\newcommand{\tcAsciinemaAllCasTargets}{113}
\newcommand{\tcAsciinemaAllCasEdges}{50}
\newcommand{\tcAsciinemaAllCasFour}{50}
\newcommand{\tcAsciinemaIntCasTargets}{86}
\newcommand{\tcAsciinemaIntCasFour}{36}
\newcommand{\tcAsciinemaRefAbsentNoGen}{34}
\newcommand{\tcAsciinemaRefObsNoGen}{145}
\newcommand{\tcAsciinemaRefKindWith}{10}
\newcommand{\tcAsciinemaRefKindProperty}{12}
\newcommand{\tcAsciinemaRefKindAttribute}{7}
\newcommand{\tcAsciinemaRefKindGenerator}{7}
\newcommand{\tcSublisterNegObserved}{19}
\newcommand{\tcSublisterMiss}{17}
\newcommand{\tcSublisterKindConstructor}{2}
\newcommand{\tcSublisterKindWith}{0}
\newcommand{\tcSublisterKindProperty}{0}
\newcommand{\tcSublisterKindClassValue}{11}
\newcommand{\tcSublisterKindAttribute}{0}
\newcommand{\tcSublisterKindParamValue}{0}
\newcommand{\tcSublisterKindSelfOverride}{1}
\newcommand{\tcSublisterKindSelfMethod}{1}
\newcommand{\tcSublisterKindViaLibrary}{2}
\newcommand{\tcSublisterKindGenerator}{0}
\newcommand{\tcSublisterLeftDispatch}{4}
\newcommand{\tcSublisterLeftImplicit}{16}
\newcommand{\tcSublisterLeftDispatchImplicit}{1}
\newcommand{\tcSublisterAllCasTargets}{19}
\newcommand{\tcSublisterAllCasEdges}{7}
\newcommand{\tcSublisterAllCasFirst}{11}
\newcommand{\tcSublisterRefObs}{87}
\newcommand{\tcSublisterRefAbsent}{21}
\newcommand{\tcAsciinemaParamGain}{23}
\newcommand{\caAsciinemaWrong}{57}
\newcommand{\caSublisterWrong}{18}
\newcommand{\jvAsciinemaEdgesOld}{406}
\newcommand{\jvAsciinemaEdgesNew}{310}
\newcommand{\jvSublisterEdgesOld}{516}
\newcommand{\jvSublisterEdgesNew}{295}
\newcommand{\toolJoernNew}{4.0.661}
\newcommand{\dynAsciinemaAllRuleNoNameDeg}{14}
\newcommand{\dynSublisterAllRuleNoNameDeg}{0}
\newcommand{\sigAsciinemaSparse}{15}
\newcommand{\sigSublisterSparse}{0}
\newcommand{\tcAsciinemaAllCasThree}{27}
\newcommand{\dynRuns}{3}
\newcommand{\toolPython}{3.8}
\newcommand{\toolCoverage}{7.2.7}
\newcommand{\kcRows}{k01 & \texttt{C()}, then \texttt{c.m()} & 2/2 & 2/2 \\ k02 & \texttt{mod.C()} & 1/1 & 1/1 \\ k03 & \texttt{with C():} & 1/3 & 1/3 \\ k04 & \texttt{for cls in [A, B]: cls(1).go()} & 2/3 & 2/3 \\ k05 & \texttt{set\_defaults(cmd=C)}; \texttt{args.cmd(a).execute()} & 2/4 & 2/4 \\ k06 & \texttt{self.api = Api()}; \texttt{self.api.upload()} & 4/4 & 3/4 \\ k07 & \texttt{a = Api()}; \texttt{a.hostname()} & 2/2 & 2/2 \\ k08 & parameter \texttt{n.notify()} & 2/2 & 2/2 \\ k09 & \texttt{@property} read & 1/2 & 1/2 \\ k10 & \texttt{self.m()}, same class & 2/2 & 2/2 \\ k11 & \texttt{table[\textquotesingle h\textquotesingle]()} & 0/1 & 0/1 \\ k12 & as k05, class imported & 0/2 & 0/2 \\ k13 & as k08, class imported & 2/2 & 2/2 \\ k14 & as k06, class imported & 3/4 & 3/4 \\ k15 & parameter, two classes with \texttt{notify}; one passed & 2/2 & 2/2 \\ k16 & \texttt{def record(p, record=mod.record)}; \texttt{record(p)} & 1/2 & 1/2 \\ k17 & as k16, default defined in the same file & 2/2 & 2/2 \\ k18 & as k03, class defined after \texttt{run} & 1/3 & 1/3 \\ k19 & as k01, class defined after \texttt{run} & 1/1 & 1/1 \\ k20 & \texttt{with C(f)} inside \texttt{with open(p) as f} & 1/3 & 1/3 \\ k21 & \texttt{with cm.C(x)}, \texttt{cm} a module alias & 1/3 & 1/3 \\ k22 & as k18, class inherits an imported class & 1/3 & 1/3 \\ k23 & as k01, class name in lower case & 0/1 & 0/1 \\ k24 & as k03, class name in lower case & 0/3 & 0/3 \\ k25 & as k23, called as \texttt{mod.c(x)} & 0/1 & 0/1 \\ k26 & as k23, class in a package \texttt{\_\_init\_\_} & 0/1 & 0/1 \\ k27 & as k24, class imported by name & 1/3 & 1/3 \\}
\newcommand{\dynAsciinemaPropExec}{12}
\newcommand{\dynAsciinemaPropKept}{12}
\newcommand{\dynSecExecutedInt}{2}
\newcommand{\tcCtorLowerMissing}{9}
\newcommand{\tcCtorLower}{11}
\newcommand{\tcCtorUpper}{7}
\newcommand{\edAsciinemaRecovered}{10}
\newcommand{\edSublisterRecovered}{0}
\newcommand{\edAsciinemaRecoveredProps}{8}
\newcommand{\tcAsciinemaAloneVia}{1}
\newcommand{\kcCases}{27}
\newcommand{\kcUnreachedOld}{17}
\newcommand{\kcUnreachedNew}{18}
\newcommand{\toolJoern}{4.0.569}
\newcommand{\toolSemgrep}{1.180.0}
\newcommand{\codeCommit}{3c1218c}
\newcommand{\nr}{\texttt{not\_reachable}}
\newcommand{\re}{\texttt{reachable}}
\newcommand{\unk}{\texttt{unknown}}
\newcommand{\app}{\textsf{APP}}
\newcommand{\lib}{\textsf{LIB}}
\newcommand{\libc}{\textsf{LIB+C}}
\newcommand{\sint}{\textsf{int}}
\newcommand{\sext}{\textsf{ext}}
\def\artifactDOI{10.5281/zenodo.23554742}
\newcommand{\ifdoi}[2]{\ifx\artifactDOI\empty#2\else#1\fi}
\newcommand{\doiurl}{\begingroup\edef\x{\endgroup\noexpand\url{https://doi.org/\artifactDOI}}\x}

\begin{document}

\title{When Static Reachability Says No:\\Checking Joern-Based Negative Verdicts for Python Against Execution}

\author{
\IEEEauthorblockN{Artur Zięba-Kozarzewski}
\IEEEauthorblockA{KRYPTON Polska Sp.\ z o.o.\\
a.zieba@krypton-polska.com\\
ORCID: 0000-0003-4696-4983}
\and
\IEEEauthorblockN{Bartosz Grzegorzewski}
\IEEEauthorblockA{KRYPTON Polska Sp.\ z o.o.\ (work done as an intern)\\
bartoszgrzegorzewski@gmail.com\\
ORCID: 0009-0002-7696-3308}
}

\maketitle

\begin{abstract}
A negative reachability verdict removes a finding from human review. We check such verdicts
against execution. Our pipeline, SourceLens, maps Semgrep findings to a Python program's call graph,
whose edges a query takes from the calls that Joern links in its code property graph. We ran two
command-line programs from the PyCG macro-benchmark, in which the benchmark's manual call graph and
Joern disagree, under coverage and recorded the calls between their functions. On these edges,
\dynAsciinemaNeg{} functions of asciinema are unreachable from the program's entry points. The project's own integration script executes
\dynAsciinemaIntNegExec{} of them, and all our runs \dynAsciinemaAllNegExec. In Sublist3r,
\dynSublisterAllNegExec{} of \dynSublisterNeg{} are executed. The latest Joern release gives the
same verdicts. The missing calls go through values, such as classes and functions passed around, or
have no call expression, as in \texttt{with} blocks and property reads. In asciinema, four missing
edges make \tcAsciinemaAllCasFour{} of the \dynAsciinemaAllNegExec{} executed functions reachable.
The manual call graphs also lack observed calls, in asciinema many of
them property reads and \texttt{with} blocks, which the benchmark may not count as calls. A rule
that turns negative verdicts into \emph{unknown} recovers \dynAsciinemaAllRuleDeg{} of the
\dynAsciinemaAllNegExec{} and \dynSublisterAllRuleDeg{} of the \dynSublisterAllNegExec, mostly
through dunder and private method names. It recovers \pjTwoDegraded{} of the \pjTwoNegBefore{}
security findings with a negative verdict, all in executed functions. Its selectivity cannot be measured,
as only \dynAsciinemaAllNotExecNeg{} and \dynSublisterAllNotExecNeg{} negative verdicts were not
executed, and under library entry-point models it degrades every negative verdict.
\end{abstract}


\section{Introduction}\label{sec:intro}

Dependency scanners match component versions against vulnerability databases and report
many vulnerabilities in code that a project never executes. Zhou et al.\ report a 92.0\%
false-positive rate for SBOM-based scanning on manually audited repositories and show that checking
whether the vulnerable functions are used, in their experiment through coverage, removes 61.9\% of
these false positives~\cite{Zhou26}. Reachability analysis is
therefore attractive as a filter. Its output, however, is not neutral. A verdict that a
finding is not reachable is an instruction to stop looking at it. A vulnerable library function
is reachable only through a path from the application's entry points to the call site that
invokes it, and that path runs through the application's own call graph.

Static call graphs are incomplete in practice~\cite{Sui20,Samhi24,Bouzenia24}, and Python makes many calls
through attributes, decorators and values passed around. To learn how often a negative verdict
is wrong, one needs a reference for the negative class. A second static call graph is a weak
reference, because it can miss the same calls. Execution is a one-sided reference. A function that
runs is reachable, so every executed function with a negative verdict is an error, and the count
is a lower bound.

We check negative verdicts on the application's own call graph. We built SourceLens, a pipeline
that maps Semgrep findings in the application's code to a Joern call graph and returns
a three-valued verdict: \re, \nr{} or \unk. A degradation rule turns a negative verdict into
\unk{} when evidence about the specific target suggests that the graph is incomplete around it.
We ran two programs of the PyCG macro-benchmark~\cite{Salis21}, asciinema
and Sublist3r, under coverage, recorded the calls between their functions, and ask three questions:

\begin{itemize}
\item \textbf{RQ1.} How many negative reachability verdicts does execution contradict?
\item \textbf{RQ2.} Which missing call edges cause the contradicted verdicts?
\item \textbf{RQ3.} How many contradicted verdicts does a degradation rule with target-local signals recover?
\end{itemize}

In both programs, most negative verdicts are contradicted. The missing edges are calls whose
callee is a value, and calls that have no call expression. Because one missing edge cuts off
everything below it, a handful of call sites decide the verdicts of dozens of functions. The
manually built call graphs of the benchmark also lack observed calls. In Sublist3r, these are calls
through \texttt{self} to overriding methods; in asciinema, many are property reads and \texttt{with}
blocks, which the benchmark may not count as calls. Used as a reference, they agree with many of the
contradicted verdicts. The rule recovers about half of the contradicted verdicts in asciinema,
mostly through dunder and private method names, none of the property reads and none of the
security findings. Because few negative verdicts remain uncontradicted, we cannot measure whether
the rule is selective.

The paper contributes:
\begin{itemize}
\item A check of Joern-based negative reachability verdicts for two Python programs against
execution, with lower bounds on the errors, and the same negative verdicts under a second Joern release.
\item An attribution of the errors to missing call edges of a few syntactic kinds, measured on
observed calls, with an ablation, the cascade of single edges, and minimal test cases on two
Joern releases.
\item An evaluation of a target-local degradation rule against execution.
\end{itemize}
\ifdoi{Scenarios, coverage data, call traces, test cases and code are available as an
artifact~\cite{ReachArtifact}.}{Scenarios, coverage data, call traces, test cases and code will be available as an artifact.}

\section{Background and Related Work}\label{sec:related}

\textbf{Reachability as a precision instrument.} Function-level reachability or usage is used to prune
dependency-vulnerability alerts~\cite{Zhou26,Patel25}. Similar pruning underlies
ecosystem-scale studies in Maven~\cite{Wu23,Mir23}. In Python, Drosos et al.\ use reachability
over stitched PyCG call graphs to classify dependency code as bloated, and note that the static
graph may contain false negatives~\cite{Drosos24}. Zhou et al.\ list reflection and dynamic code
loading among the threats to their analysis~\cite{Zhou26}, and Mir et al.\ argue that false
negatives cost more than false positives in security analysis~\cite{Mir23}.

\textbf{Measuring what reachability misses.} Eclipse Steady combines static analysis with
test-driven dynamic analysis. In 7.9\% and later 11.7\% of the cases with reachable vulnerable
code, reachability was established only by the combination of techniques, with reflection
and inversion of control named as causes~\cite{Ponta18,Ponta20}. Imtiaz et al.\ found 11
alerts on OpenMRS whose vulnerable code was executed by unit tests, while static analysis
deemed the dependency unused~\cite{Imtiaz21}. Chuang et al.\ show for a large industrial Java
project that adding call edges for dynamic features turns part of the ``unused'' dependency
verdicts into used ones, and argue against a binary reachable/unreachable verdict~\cite{Chuang22}.
S\={o}jiTantei reports a 0.476 false-negative rate against a manual baseline on npm
clients~\cite{Chinthanet22}. ConVReach evaluates dependency reachability in C/C++ against
manually labelled cases~\cite{Feng26}, and Patel et al.\ compare sound and unsound binary call
graphs with a compiler-level one~\cite{Patel25}. Test-generation approaches establish that a library
vulnerability can be triggered from client code~\cite{Chen24}. Like Eclipse Steady and Imtiaz et al., we
use execution to audit negative verdicts. Execution also serves as a reference for call graphs.
Sui et al.\ and Helm et al.\ build dynamic baselines from tests or fixed inputs~\cite{Sui20,Helm24},
and JARVIS builds its Python reference from test traces and manual inspection~\cite{Huang23}.
DyPyBench compares static and dynamic call graphs of executable Python projects~\cite{Bouzenia24}.
We use such a reference for the negative verdicts of a reachability filter and for the call
edges whose absence produces them.

\textbf{Call-graph soundness.} Sui et al.\ measure a median recall of 0.884 on 31 Java
programs against a dynamic oracle~\cite{Sui20}. On Android, static tools miss 61\% of the
dynamically executed methods, and constructors are prominent among the misses~\cite{Samhi24}.
In Java analysis frameworks, missing call-graph elements in the core analysis stage most often stem
from mishandled implicit invocations of special methods~\cite{Zhang24}. For Python, PyCG reports
about 70\% average recall against manually built call graphs~\cite{Salis21}. Against dynamic call
graphs of 39 executable projects, only 49\% of the observed edges and 48.5\% of the observed callees
appear in PyCG's graphs, mostly because calls on built-in types are not modelled~\cite{Bouzenia24}.
JARVIS misses calls to functions stored in lists, tuples and dictionaries, and its authors note that
PyCG does not handle \texttt{with} statements~\cite{Huang23}. A points-to analysis for Python
finds edges that PyCG misses, and its authors conclude that recall rather than precision is the
more pressing issue~\cite{RakAmnouykit25}. A single unresolved dynamic
construct can cut off large parts of a program. Unresolved reflection can leave a Java call graph
empty~\cite{Helm24}, and dynamically computed module names sometimes drop entire JavaScript
modules~\cite{Laursen24}. Without runtime heap information, static analysis often reaches less
than 10\% of some Java benchmarks~\cite{Grech17}. Chakraborty et al.\ attribute missing edges
in JavaScript to root causes from dynamic data flows, including inter-dependent calls, and find
that the causes vary across programs~\cite{Chakraborty22}. A call through a function passed as a
parameter can only be resolved once the calls that supply it are inferred, and one additional edge
can change overall reachability markedly~\cite{Chakraborty24}. Recovering dynamic property accesses
at the point of dispatch raises recall~\cite{Laursen24}. Our cascade results in
Section~\ref{sec:rq2} show an analogous dependence for reachability in Python.

\textbf{Entry-point models.} Which code is reachable depends on where the analysis starts. Reif
et al.\ show that a call graph for a library has to follow the purpose of the client analysis,
and describe the common practice of treating all non-private methods as entry
points~\cite{Reif16}. Ali and Lhoták build call graphs of an application alone under a separate
compilation assumption. The library can call back only into application methods whose names it
knows, and reflective access to application internals is assumed to be rare~\cite{Ali12}. Helm et al.\ name the main methods
of applications and the public API of libraries as typical choices~\cite{Helm24}. Huang et al.\
report a recall of 0.76 for PyCG when every function is an entry point. When accuracy is computed
along call chains from the application's entry functions, the recall is 0.61, and they attribute
the drop mainly to this difference in how accuracy is computed~\cite{Huang23}. We study programs from their own
entry point and use library models only to show a structural limit of the rule.

\textbf{Soundiness.} Livshits et al.\ argue that practical whole-program
analyses are unsound by design, and ask evaluations to compare results with observed dynamic
behaviour and to report the consequences of unsoundness for the analysed
programs~\cite{Livshits15}.

\section{Approach}\label{sec:approach}

\textbf{Pipeline.} SourceLens runs Semgrep on a source tree, classifies each finding by scope
(application, test, build, documentation) and maps it to its enclosing function. It builds one
Joern code property graph~\cite{Yamaguchi14} for the whole tree, with the default options of
Joern's Python frontend. A call edge links a method to each callee of its call sites that Joern
resolves to a method of the analysed code (\texttt{m.call.callee}, external methods removed). A
graph search from a set of entry points decides whether the enclosing function is reachable and
reconstructs the call chain. In the application model \app, the entry points are the module code of
every file and the functions named \texttt{main}. This set contains the program's real entry point
and more, so it can only make negative verdicts rarer. The verdicts thus depend both on the graph
Joern builds and on this edge definition (Section~\ref{sec:rq2} separates the two); for brevity we
call the verdicts computed on these edges Joern's verdicts.

\textbf{Three-valued verdict.} A \nr{} verdict silences a finding. An \unk{} verdict states
that the graph does not support a conclusion and is never silenced. The degradation rule is
the only source of \unk{} in this study.

\textbf{Degradation rule.} The rule inspects each \nr{} verdict and turns it into \unk{} if
any of five signals fires (Table~\ref{tab:signals}). Four signals concern the target. They fire
on an implicitly invoked or private method and on a class-level target whose methods are all
unreached or whose name suggests a framework role. They also fire on a target with no resolved
edge anywhere in the graph and on a sparse neighbourhood in the target's own file. The fifth is a floor for a
graph that has almost no edges. Since four of five signals are target-local, we call the rule
target-local. Positive verdicts are never degraded. The signals were derived
from the two CVE-Bench misses described in Section~\ref{sec:rq2} and from an earlier run with an
empty graph. The rule carries to call graphs an idea from our earlier work on declared dependency
graphs~\cite{Topo26}. There, a missing path is reported as unknown rather than used as a veto,
and a graph-completeness signal decides when a missing path cannot be trusted.

\begin{table}[t]
\caption{Degradation signals. Any signal turns \nr{} into \unk{}; \re{} is never changed.}
\label{tab:signals}
\centering
\footnotesize
\begin{tabular}{@{}p{0.34\columnwidth}p{0.58\columnwidth}@{}}
\toprule
Signal & Condition \\
\midrule
Implicit or private target & method name is a dunder (e.g.\ \texttt{\_\_init\_\_}) or starts with an underscore \\
Class-level or framework role & target is a class with no method reached, or its name contains a framework hint (e.g.\ \emph{authenticator}, \emph{plugin}, \emph{adapter}) \\
Isolated target & target has no resolved incoming or outgoing edge in the whole graph \\
Sparse neighbourhood & resolved edges per method in the target's file below \ruleLocalSparse{} \\
Global floor & edges per method in the whole graph below \ruleGlobalFloor{} \\
\bottomrule
\end{tabular}
\end{table}

\section{Study Design}\label{sec:design}

\textbf{Programs.} The PyCG macro-benchmark~\cite{Salis21} provides \calProjects{} Python projects
with manually built call graphs. We use repository commit \texttt{\calTruthCommit} and the project
commits it pins. The \app{} model fits a project only if its program entry point is in the
analysed code. asciinema and Sublist3r have a \texttt{main}. face\_classification consists of
scripts. Its \appFaceClsNeg{} negative verdicts all agree with the manual graph, and we did not
execute it, because its scripts need trained models and datasets. fabric's entry point is a
console script declared in package metadata, and under \app{} the manual graph reaches
\fabAppManualReached{} of its \fabAppPop{} functions. autojump's entry script has no
\texttt{.py} suffix and is absent from both graphs. \app{} thus fits three projects, and we executed
two of them. We fixed these criteria in an earlier analysis
of the same release run, after its first results. We study asciinema and Sublist3r. In the three other
projects, the manual graph agrees with Joern on every negative verdict, reaches almost no code, or
lacks the entry script. Our counts therefore describe programs in which the two graphs disagree,
not a sample of Python programs.

\textbf{Verdicts.} The verdicts come from the release run of SourceLens: one run of version 0.9
at code commit \texttt{\codeCommit} with Joern \toolJoern{} and Semgrep \toolSemgrep{} on all
projects and on CVE-Bench, with every input pinned. The population consists of the source-level
functions in files of application scope, without Joern's synthetic methods: \popAsciinema{} in
asciinema and \popSublister{} in Sublist3r. A function has a negative verdict if no path in the
saved Joern graph leads to it from the \app{} entry points. Joern methods are matched to source
functions by qualified name.

\textbf{Execution.} We ran both programs in a Python \toolPython{} container without network
access, with coverage.py \toolCoverage{} active in every process, including forked worker
processes. A function counts as \emph{executed} if at least one statement of its own body ran,
excluding nested definitions, a leading docstring and the \texttt{def} line, which runs at import. Two scenario
sets drive asciinema. \sint{} is the project's integration script \texttt{tests/integration.sh},
unchanged: help and version, \texttt{auth}, \texttt{play} and \texttt{cat} on recordings in both
formats, and \texttt{rec} with signals, stdin capture, raw output and appending. \sext{} is our
own set. It adds uploads to a local stand-in for the asciinema API, playback from local URLs and
web pages, error paths, settings from the configuration file, and notifications triggered by the
pause key during recording. We wrote \sext{} after seeing the verdicts, and some scenarios target
known findings. Sublist3r has no scripted runs. Our scenarios follow its README and cover the
remaining command-line options. Without network, its search requests fail, and the code that
processes successful responses does not run. We ran all scenarios \dynRuns{} times. The sets of
executed functions were identical, and we report the first run. Executed lines differed in one
signal-handling branch.

\textbf{Observed calls.} In the third run, a profiling hook recorded each call between two project
functions. The caller is the innermost project frame that encloses the call. If library frames lie in
between, as for a thread target or an argparse type converter, the call is marked as made through
a library. When a generator resumes, the hook sees the consumer as the caller, not the function
that created the generator.

\textbf{Kinds of missing edges.} An observed call that has no edge in the Joern graph is a
\emph{missing edge}. A script classifies each by rules over the syntax tree of the caller and
assigns the first matching kind in a fixed order.
\emph{Implicit} calls have no call expression for the callee: \texttt{C(\ldots)} reaching
\texttt{C.\_\_init\_\_}, \texttt{with} reaching \texttt{\_\_enter\_\_} and \texttt{\_\_exit\_\_},
and the read of a \texttt{@property}. In a \emph{value} call, a value determines the callee. This covers a
class or function passed around and called later, a method called on an object held in an
attribute, variable or parameter, and \texttt{self.m()} reaching an override in a subclass. A
\emph{static} call names its callee through a plain name, \texttt{self}, \texttt{super()} or a
module alias. Calls made through a library and generator resumptions form their own groups. Unlike
the JARVIS reference, which filters out some interpreter-invoked calls~\cite{Huang23}, we keep
implicit calls between project functions. The rules are syntactic and were not checked by an
independent coder.

\textbf{Ablation and cascade.} We add the missing edges of one or more groups to the Joern graph
and count the executed functions with a negative verdict that become reachable. Because an edge
helps only once its caller is reachable, the groups do not add up, and we report every
combination. For the cascade, we pick missing edges one at a time, each time the edge that makes
the most remaining executed functions reachable.

\textbf{Manual call graphs.} The PyCG authors built the call graphs of the macro-benchmark by
hand, spending on average 10 hours per project, to measure the precision and recall of call
edges~\cite{Salis21}. We compare the observed direct calls with them. We did not establish
whether property reads and context-manager calls fall under the benchmark's notion of a call.

\textbf{Minimal test cases.} We wrote \kcCases{} small modules for call patterns seen in the missing
edges, each with a function \texttt{run} that calls the expected callees when executed. We built
their graphs with Joern \toolJoern{} and with \toolJoernNew, the latest release on 20 October
2026, using the edge definition of the pipeline. We also extracted the method references of the
graph, to test whether a query that follows them would link more calls.

\textbf{Reproducibility.} Each analysis first reproduces all \repPycgVerdicts{} verdicts and
\repPycgDegradations{} degradation decisions of the release run on the projects and all
\repCbVerdicts{} verdicts on CVE-Bench. We extracted the graphs of both programs again with Joern
\toolJoern. The call edges were identical to the release run (\jvAsciinemaEdgesOld{} and
\jvSublisterEdgesOld{} distinct edges). Joern \toolJoernNew{} produced different graphs
(\jvAsciinemaEdgesNew{} and \jvSublisterEdgesNew{} edges) with the same set of negative verdicts in
both programs. Missing edges, ablation and cascade are computed on the \toolJoern{} graphs; for
\toolJoernNew{} we compare the negative verdicts and the test cases.

\section{Results}\label{sec:results}

\subsection{RQ1: How many negative verdicts does execution contradict?}\label{sec:rq1}

Table~\ref{tab:exec} compares the verdicts with execution. In asciinema, Joern reaches
\dynAsciinemaPos{} of the \popAsciinema{} functions from the entry points. The integration
script alone executes \dynAsciinemaIntNegExec{} of the \dynAsciinemaNeg{} others, and all scenarios
together \dynAsciinemaAllNegExec. In Sublist3r, \dynSublisterAllNegExec{} of \dynSublisterNeg{}
negative verdicts are contradicted. These counts are lower bounds. A function that did not run in
our scenarios may still be reachable.

\begin{table}[t]
\caption{Negative verdicts against execution under \app. Joern neg: functions Joern does not reach;
manual: verdicts of the manual call graph on the same functions. Cells: executed / total. For
Sublist3r, which has no scripted runs, ``all'' consists of our scenarios only.}
\label{tab:exec}
\centering
\footnotesize\setlength{\tabcolsep}{3.5pt}
\begin{tabular}{@{}lrrrr@{}}
\toprule
 & \multicolumn{3}{c}{asciinema} & Sublist3r \\
\cmidrule(lr){2-4}\cmidrule(l){5-5}
 & \sint & \sext & all & all \\
\midrule
functions executed & \dynAsciinemaIntExec/\popAsciinema & \dynAsciinemaExtExec/\popAsciinema & \dynAsciinemaAllExec/\popAsciinema & \dynSublisterAllExec/\popSublister \\
Joern neg & \dynAsciinemaIntNegExec/\dynAsciinemaNeg & \dynAsciinemaExtNegExec/\dynAsciinemaNeg & \dynAsciinemaAllNegExec/\dynAsciinemaNeg & \dynSublisterAllNegExec/\dynSublisterNeg \\
\quad manual reaches & \dynAsciinemaIntWrongExec/\dynAsciinemaWrong & \dynAsciinemaExtWrongExec/\dynAsciinemaWrong & \dynAsciinemaAllWrongExec/\dynAsciinemaWrong & \dynSublisterAllWrongExec/\dynSublisterWrong \\
\quad manual does not & \dynAsciinemaIntAgreeExec/\dynAsciinemaAgree & \dynAsciinemaExtAgreeExec/\dynAsciinemaAgree & \dynAsciinemaAllAgreeExec/\dynAsciinemaAgree & \dynSublisterAllAgreeExec/\dynSublisterAgree \\
manual neg & \dynAsciinemaIntRefNegExec/\dynAsciinemaRefNeg & \dynAsciinemaExtRefNegExec/\dynAsciinemaRefNeg & \dynAsciinemaAllRefNegExec/\dynAsciinemaRefNeg & \dynSublisterAllRefNegExec/\dynSublisterRefNeg \\
\bottomrule
\end{tabular}
\end{table}

\textbf{The manual call graphs.} Of the functions that Joern does not reach in asciinema, the
manual graph reaches \dynAsciinemaWrong. The remaining \dynAsciinemaAgree{} would count as correct
verdicts against the manual graph, and \dynAsciinemaAllAgreeExec{} of these \dynAsciinemaAgree{} are
executed. In
Sublist3r, Joern reaches \dynSublisterJoernOnly{} functions that the manual graph does not reach,
and \dynSublisterAllJoernOnlyExec{} of them are executed. The reason is visible in the observed
calls. Of the distinct direct calls observed in asciinema between functions of the manual graph,
\tcAsciinemaRefAbsentNoGen{} of \tcAsciinemaRefObsNoGen{} have no edge in it, not counting the
\tcAsciinemaRefKindGenerator{} absent generator resumptions.
They include \tcAsciinemaRefKindProperty{} property reads, \tcAsciinemaRefKindWith{} calls from
\texttt{with} blocks and \tcAsciinemaRefKindAttribute{} calls on objects held in attributes.
Whether the benchmark counts property reads and context-manager calls as calls is open
(Section~\ref{sec:design}), so only the remaining absent calls of asciinema are omissions under any
reading. In Sublist3r, \tcSublisterRefAbsent{} of \tcSublisterRefObs{} observed direct calls are
absent, all of them calls through \texttt{self} that reach overriding methods of the search-engine
classes, which have a call expression.

\textbf{Security findings.} Table~\ref{tab:projects} shows the findings of security rules. In the two programs, \pjTwoNegAfter{}
of \pjTwoFindings{} keep a \nr{} verdict after degradation, all in
asciinema. We traced them by hand in the source (Appendix~\ref{app:chains}). All
\dynSecTotal{} are executed in our runs, \dynSecExecutedInt{} of them by the integration script
alone. The other three need our targeted \sext{} scenarios. The manual graph reaches
\survRefWrong{} of them.

\begin{table}[t]
\caption{Verdicts on findings of security rules in the release run. The release run covered all
\calProjects{} projects; \app{} fits three of them, and we study two (asciinema and Sublist3r).
\re{} without a chain: the finding lies in an entry point, such as module code.}
\label{tab:projects}
\centering
\footnotesize
\begin{tabular}{@{}lrr@{}}
\toprule
 & studied & all \calProjects{} \\
\midrule
Findings & \pjTwoFindings & \pjSecFindings \\
\re{} with derived call chain & \pjTwoChain & \pjSecChain \\
\re{}, finding in an entry point & \pjTwoEntry & \pjSecEntry \\
\nr{} before degradation & \pjTwoNegBefore & \pjSecNegBefore \\
degraded to \unk & \pjTwoDegraded & \pjSecDegraded \\
\nr{} after degradation & \pjTwoNegAfter & \pjSecNegAfter \\
\bottomrule
\end{tabular}
\end{table}

\subsection{RQ2: Which missing call edges cause the contradicted verdicts?}\label{sec:rq2}

\textbf{Kinds.} Table~\ref{tab:kinds} lists the missing edges by kind. In asciinema, the most
frequent are calls on objects held in attributes or parameters, \texttt{with} blocks and property
reads. In Sublist3r, most missing edges come from one construct. \texttt{main} builds a list of
search-engine classes and instantiates each element, so the class is a value.

\begin{table}[t]
\caption{Observed calls without an edge in the Joern graph, by syntax of the call site.}
\label{tab:kinds}
\centering
\footnotesize
\begin{tabular}{@{}llrr@{}}
\toprule
Group & Kind & asciinema & Sublist3r \\
\midrule
implicit & \texttt{C(\ldots)} to \texttt{\_\_init\_\_} & \tcAsciinemaKindConstructor & \tcSublisterKindConstructor \\
 & \texttt{with} & \tcAsciinemaKindWith & \tcSublisterKindWith \\
 & property read & \tcAsciinemaKindProperty & \tcSublisterKindProperty \\
value & class passed as a value & \tcAsciinemaKindClassValue & \tcSublisterKindClassValue \\
 & method on an object & \tcAsciinemaKindAttribute & \tcSublisterKindAttribute \\
 & function passed as a parameter & \tcAsciinemaKindParamValue & \tcSublisterKindParamValue \\
 & \texttt{self.m()} to an override & \tcAsciinemaKindSelfOverride & \tcSublisterKindSelfOverride \\
static & \texttt{self.m()}, same class & \tcAsciinemaKindSelfMethod & \tcSublisterKindSelfMethod \\
library & through library frames & \tcAsciinemaKindViaLibrary & \tcSublisterKindViaLibrary \\
other & generator resumption & \tcAsciinemaKindGenerator & \tcSublisterKindGenerator \\
\midrule
 & total & \tcAsciinemaMiss & \tcSublisterMiss \\
\bottomrule
\end{tabular}
\end{table}

\textbf{Ablation.} In asciinema, \tcAsciinemaNegObserved{} functions with a negative verdict are
observed as callees. Adding the value edges leaves \tcAsciinemaLeftDispatch{} of them unreachable,
adding the implicit edges leaves \tcAsciinemaLeftImplicit, and adding both leaves
\tcAsciinemaLeftDispatchImplicit. The edges through library frames close the rest. The causes do
not add up. \tcAsciinemaAloneDispatch{} functions become reachable through value edges alone,
\tcAsciinemaAloneImplicit{} through implicit edges alone, \tcAsciinemaAloneVia{} through an edge
made through a library, and \tcAsciinemaOnlyJointly{} only when edges of several groups are present. In Sublist3r, value edges leave \tcSublisterLeftDispatch{}
of \tcSublisterNegObserved{} unreachable, implicit edges \tcSublisterLeftImplicit, and both
\tcSublisterLeftDispatchImplicit. One missed call through \texttt{self} within a class closes the
rest. Against the manual graph as the oracle, the same two groups are also sufficient: they
explain all \caAsciinemaWrong{} and \caSublisterWrong{} disagreements.

\textbf{Cascade.} A few edges carry a large share of the effect. In asciinema, the command-line entry point
registers each subcommand with \texttt{set\_defaults(cmd=\ldots)}, later creates the command
with \texttt{args.cmd(\ldots)} and calls its \texttt{execute()}. The first three edges of the cascade restore this dispatch
for playback and recording and make \tcAsciinemaAllCasThree{} executed functions reachable. The
fourth is \texttt{recorder.record} calling \texttt{record(\ldots)}, a parameter whose default value
is \texttt{pty.record}. It adds \tcAsciinemaParamGain. These four edges make
\tcAsciinemaAllCasFour{} of the \tcAsciinemaAllCasTargets{} functions reachable. Reaching every one of
the \tcAsciinemaAllCasTargets{} takes \tcAsciinemaAllCasEdges{} edges in total, most of which add a
single function. With the
integration script alone, the same four edges make \tcAsciinemaIntCasFour{} of
\tcAsciinemaIntCasTargets{} functions reachable. In Sublist3r, the first edge, from \texttt{main}
to one search-engine class, makes \tcSublisterAllCasFirst{} of \tcSublisterAllCasTargets{}
reachable, and \tcSublisterAllCasEdges{} edges reach all of them. How much a missing edge matters
depends on the code below it, not on how often its kind occurs.

\textbf{Minimal test cases.} Table~\ref{tab:cases} shows which expected callees Joern reaches
in the test cases. Under the pipeline's edge definition, both releases miss \texttt{with}, property
reads, and classes and functions used as values. They also miss a parameter whose default is a
function of another module (k16) and methods on attributes that hold an object of an imported class. With the default
function in the same file (k17), the call is linked. Constructor calls are linked when the class
name starts with an upper-case letter (k01, k02, k18--k22), also inside \texttt{with}. They are not
linked when it starts with a lower-case letter and the class is called by name in its own module or
through a module or package alias (k23--k26). In both programs, \tcCtorLowerMissing{} of the
\tcCtorLower{} observed constructor calls of lower-case classes have no edge, and all
\tcCtorUpper{} of upper-case classes have one. Each missed call names the class in its own module
or through a module name. The linked lower-case calls all
construct \texttt{raw} in asciinema, which the calling modules import by name
(\texttt{from asciinema.term import raw}). The test case for this form (k27) is linked on both
releases, so the lower-case calls are missed in exactly the forms of k23--k26. In \kcUnreachedOld{} of \kcCases{} cases with \toolJoern{} and
\kcUnreachedNew{} with \toolJoernNew, at least one expected callee is not reached; the newer release
also loses a method on an attribute in a single file (k06). Some calls on receivers of unknown type
are linked to every method of the same name: in k15, \texttt{N2.notify} is reached although only
\texttt{N1} is passed. This is consistent with Joern reaching functions that the manual graph does
not. k10 passes, yet one call through \texttt{self} within a class is missing in Sublist3r; the
test cases do not isolate what differs there. The open issue \#6413\footnote{\url{https://github.com/joernio/joern/issues/6413} and
\url{https://github.com/joernio/joern/issues/5357}, accessed 20 October 2026.} of the Joern repository reports parameter default values missing
from the graph, the pattern behind \texttt{record=pty.record}; issue \#5357 reports shortcomings
in handling Python imports.

\textbf{Edge definition.} The pipeline links calls only (\texttt{m.call.callee}). The graph also
holds method references. In the test cases, a \texttt{with} block on an upper-case class of the same
module carries references to \texttt{\_\_enter\_\_} and \texttt{\_\_exit\_\_} (k03, k18, k20), while
functions and classes used as values appear as plain identifiers. In the two programs, no
\texttt{with} block carries such a reference. Following all method references inside method bodies
makes \edAsciinemaRecovered{} more of the \dynAsciinemaNeg{} functions of asciinema reachable,
\edAsciinemaRecoveredProps{} of them property getters, and \edSublisterRecovered{} in Sublist3r,
with the same result on both releases. The edge definition thus explains a small part of the
errors.

\begin{table}[t]
\caption{Minimal test cases: expected callees reached from \texttt{run} in the Joern graph.}
\label{tab:cases}
\centering
\footnotesize\setlength{\tabcolsep}{4pt}
\begin{tabular}{@{}lp{0.50\columnwidth}rr@{}}
\toprule
Case & Pattern & \toolJoern & \toolJoernNew \\
\midrule
\kcRows
\bottomrule
\end{tabular}
\end{table}

\textbf{A second setting.} CVE-Bench\footnote{\url{https://github.com/GiovanniGatti/cve-bench}, commit
\texttt{\cbCommit}. A collection of vulnerabilities in Python packages, each with a vulnerable and
a fixed commit and a security test, published as a repository rather than a peer-reviewed paper.}
is analysed as a library, with every public function as an entry point (model \lib). In
\cbTauto{} of its \cbTasks{} tasks the target is itself an entry point. Of the other
\cbInformative, \cbMissed{} are \nr. Both misses are implicit calls. In GHSA-r758-8hxw-4845 (justhtml), the target is
\texttt{JustHTML.\_\_init\_\_}, called by \texttt{JustHTML(\ldots)} from outside the package. In
CVE-2026-33175 (oauthenticator), the target class defines only private methods decorated with
\texttt{@default}, which the traitlets framework calls when an attribute is read before it is set.

\subsection{RQ3: How many contradicted verdicts does the rule recover?}\label{sec:rq3}

\textbf{Application model.} The rule turns \dynAsciinemaAllRuleDeg{} of the
\dynAsciinemaAllNegExec{} executed functions in asciinema into \unk{} and keeps
\dynAsciinemaAllRuleKept{} as \nr. In Sublist3r it degrades \dynSublisterAllRuleDeg{} of
\dynSublisterAllNegExec{} and keeps \dynSublisterAllRuleKept. Without the name signal it degrades
\dynAsciinemaAllRuleNoNameDeg{} and \dynSublisterAllRuleNoNameDeg. The rule thus recovers
mostly methods with dunder or private names. Property reads are implicit calls too, but the
\dynAsciinemaPropExec{} executed property getters of asciinema have public names, and the rule keeps
\dynAsciinemaPropKept{} of them as \nr. The functions behind the command dispatch
in asciinema have ordinary public names, resolved edges elsewhere and files that are not sparse.
The missing edge lies several calls above them, where a signal computed from the target and its
file cannot observe it. The global floor, the counterpart of the completeness signal for declared dependency
graphs~\cite{Topo26}, fires on no negative verdict in either program. Its file-level variant, the
sparse-neighbourhood signal, fires on \sigAsciinemaSparse{} of the
\dynAsciinemaNeg{} in asciinema and on \sigSublisterSparse{} of the \dynSublisterNeg{} in
Sublist3r. The missing edges here sit at a few call sites (Section~\ref{sec:rq2}). Since most negative verdicts are contradicted, we cannot measure
whether the rule separates wrong verdicts from correct ones. Only \dynAsciinemaAllNotExecNeg{} and
\dynSublisterAllNotExecNeg{} negative verdicts were not executed (the rule degrades
\dynAsciinemaAllNotExecDeg{} and \dynSublisterAllNotExecDeg{} of them), and not running is no proof
of unreachability.
Over-approximating call candidates, for example by method name, would reach
code cut off above the target at a cost in precision that we did not measure.

\textbf{Library models.} Under \lib{} every public function is an entry point, and under \libc{}
also the constructors of public classes. A function can then be unreachable only if its name
starts with an underscore, so the name signal fires on every negative verdict. The rule degrades
all \libTotalNeg{} negative verdicts of face\_classification, asciinema, Sublist3r and fabric
under \lib{}, all \libcTotalNeg{} under \libc{}, and \cbLibcDeg{} of \cbLibcNeg{} on the CVE-Bench
graphs under \libc. autojump is left out of all models, because its entry script is absent from
both graphs. In these models the
rule cannot be selective by construction.

\section{Threats to Validity}\label{sec:threats}

\textbf{Scope.} We execute two small command-line programs, analysed by one frontend of one
call-graph builder. The evaluations of Joern on Python that we found concern taint
analysis~\cite{BakerEffendi25,Wang26yasa}, not call edges. The counts describe these programs. The share of contradicted verdicts depends
on where in a program the first missing edge lies, as the cascade shows.

\textbf{Scenarios.} Execution proves reachability, possibly under a particular configuration or
input, such as an installed notification tool or the pause key. We consider such functions
reachable, because a negative verdict claims unreachability for every input. Our \sext{}
scenarios were written after seeing the verdicts and target some known findings. The \sint{} set
is the project's own and gives the more conservative count. Without network, Sublist3r runs its
error paths, and its counts are lower than with network access.

\textbf{Call tracing and classification.} The lower bounds rest on executed functions. Observed
edges also depend on how the caller is attributed: it is the innermost project frame, and
generator resumptions are attributed to the consumer. The kinds are syntactic. They describe how a
call is written, not how hard it is to resolve: a method on an attribute of a known class is in
the value group, as is a class chosen at run time. The classification rules were written by us and
not checked by an independent coder. The cascade is greedy and depends on the observed calls.
Verdicts depend on the query that defines call edges. Following method references as well
makes \edAsciinemaRecovered{} of \dynAsciinemaNeg{} and \edSublisterRecovered{} of
\dynSublisterNeg{} negative verdicts reachable (Section~\ref{sec:rq2}); other queries could change more.

\textbf{Test cases.} We wrote the test cases after observing the missing edges. They show that
the patterns occur, not how common they are. The explanation of name-based linking is inferred
from the behaviour of the graphs, not from Joern's code.

\textbf{Selection of programs.} We fixed the criteria for applications and the file exclusions in
an earlier analysis of the same data, after its first results. face\_classification was not
executed.

\textbf{Rule design and own tool.} The signals were derived from the two CVE-Bench misses that the
rule recovers. Two earlier versions of the rule were corrected after re-running it on the pipeline
results, which include these projects. Both corrections removed implementation errors that made a
signal fire on every input. We keep the thresholds of the release run. In an earlier analysis
against the manual graphs, we varied the sparse-neighbourhood ratio between \sensLsrMin{} and
\sensLsrMax. The difference between the degraded shares of disagreeing and agreeing verdicts in
asciinema then moved between \sensAsciiDiffMin{} and \sensAsciiDiffMax{} percentage points, and
Sublist3r did not change. Varying the global floor between \sensGefMin{} and \sensGefMax{} changed
nothing in this range.
The authors built and evaluate SourceLens. The Semgrep rules of the public repository are licensed
for internal use only, so the artifact references them by commit instead of including them.

\section{Conclusion}\label{sec:conclusion}

We studied two Python programs in which the benchmark's manual call graph and Joern disagree. In
both, most functions that our query finds unreachable in Joern's graph run when the program runs. The errors come from call edges whose callee is a value and from calls without a call
expression, and a few such edges decide the verdicts of a large share of the functions. Following
method references as well makes \edAsciinemaRecovered{} of the \dynAsciinemaNeg{} negative verdicts
of asciinema and \edSublisterRecovered{} of the \dynSublisterNeg{} of Sublist3r reachable. The PyCG manual call graphs also lack observed
calls and agree with many of these wrong verdicts. In asciinema, many of the calls they lack are
property reads and \texttt{with} blocks, which the benchmark may not count as calls. A target-local
rule recovers methods with dunder and private names, and misses property reads, the code cut off
above the target and every security finding with a negative verdict. Whether it separates wrong
from correct negative verdicts remains unmeasured, and under library entry-point models it degrades
all of them. Completeness signals computed over the whole graph or a file do not flag these
errors, because they sit at a few call sites. A negative verdict is therefore absence of evidence, and it is weakest for code behind a
dispatch. Evaluations of such verdicts need a reference that does not share the analysed graph's
blind spots, and execution is one. Resolving value-passing idioms such as argparse subcommands and
parameter defaults would change many verdicts at once. Constructor calls of lower-case
classes are a narrower pattern, which our test cases isolate: they are missed unless the class is
imported by name.

\appendices
\section{Call chains of the surviving misses in asciinema}\label{app:chains}
{\raggedright

Source: asciinema at commit \texttt{350d5e9}, the commit pinned by the PyCG macro-benchmark.
Line numbers refer to that commit. All five functions are executed in our runs.

\begin{itemize}
\item \textbf{Dispatch (all chains).} \texttt{\_\_main\_\_.py:79} registers
\texttt{set\_defaults(cmd=RecordCommand)} (likewise for play, cat, upload and auth);
\texttt{\_\_main\_\_.py:106--107} assigns \texttt{command = args.cmd(\ldots)} and calls
\texttt{command.execute()}.
\item \textbf{\texttt{pty.record}} (CWE-78, finding at \texttt{pty.py:135}).
\texttt{commands/record.py:71} calls \texttt{recorder.record(\ldots)};
\texttt{recorder.py:12} declares the default parameter \texttt{record=pty.record};
\texttt{recorder.py:49} calls \texttt{record(\ldots)}.
\item \textbf{\texttt{URLLibHttpAdapter.post}} (CWE-939, finding at
\texttt{urllib\_http\_adapter.py:76}). \texttt{commands/upload.py:13} and
\texttt{commands/record.py:101} call \texttt{self.api.upload\_asciicast(\ldots)};
\texttt{api.py:21} assigns \texttt{self.http\_adapter = \ldots URLLibHttpAdapter()};
\texttt{api.py:35} calls \texttt{self.http\_adapter.post(\ldots)}.
\item \textbf{The two \texttt{notify} methods} of \texttt{Notifier} and
\texttt{CustomCommandNotifier} (CWE-78, findings at \texttt{notifier.py:11} and \texttt{:64}). \texttt{commands/record.py:81} passes
the notifier to \texttt{recorder.record}; \texttt{recorder.py:48} wraps it in
\texttt{async\_notifier}; \texttt{pty.py:25} calls \texttt{notifier.notify(\ldots)}, which
enqueues; the worker loop at \texttt{async\_worker.py:31} calls \texttt{perform}, and
\texttt{recorder.py:95} calls \texttt{self.notifier.notify(\ldots)}.
\item \textbf{\texttt{asciicast.open\_url}} (CWE-939, finding at
\texttt{asciicast/\_\_init\_\_.py:48}). \texttt{commands/play.py:17} and
\texttt{commands/cat.py:15} enter \texttt{with asciicast.open\_from\_url(\ldots)};
\texttt{asciicast/\_\_init\_\_.py:91} in \texttt{\_\_enter\_\_} calls \texttt{open\_url(\ldots)}.
\end{itemize}

}

\section*{Artifact}
The code at commit \texttt{\codeCommit}, the pinned inputs, all results of the release run, the
scenarios, coverage data, call traces and test cases \ifdoi{are deposited at
\doiurl~\cite{ReachArtifact}.}{will be deposited with a DOI.}

\section*{Declaration of Generative AI and AI-Assisted Technologies}
During the preparation of this work the authors used Claude (Anthropic), via the Claude Code CLI
(2026), for four purposes: assisting with drafting and language of the manuscript, supporting the
literature search, helping implement the analysis tooling described in Section~\ref{sec:design}
(the execution and call-tracing harness, the analysis scripts and the syntactic classifier of
missing edges), and writing the minimal test cases. Literature identified with AI assistance was
verified against primary sources before citation. Every number of this study is generated by a
script from result files that \ifdoi{are}{will be} part of the artifact. The authors reviewed and edited all content and take
full responsibility for the content of the published article. The generative-AI tool is not, and
cannot be, an author. A complete record of the assisted workflow is preserved in the project's
version-control history.

\bibliographystyle{IEEEtran}
\bibliography{refs}

\end{document}